\documentclass{appolb}
\usepackage{epsfig,dcolumn}
\usepackage{graphicx}
\usepackage{color}

\usepackage{amsmath}
\usepackage{amsfonts}
\usepackage{amssymb}
\usepackage{graphicx}
\usepackage{type1cm}
\usepackage{eso-pic}

\newcommand{\be}{\begin{equation}}
\newcommand{\ee}{\end{equation}}

\newcommand{\im}{\mathrm{Im}\,}
\newcommand{\re}{\mathrm{Re}\,}

\usepackage{bm} 

\begin{document}

\title{Analytic approach to $\pi K$ scattering and strange resonances%
\thanks{Presented at Excited QCD 2017, Sintra, Portugal, May 7-13}%
}
\author{A.Rodas
\address{Departamento de F\'isica Te\'orica II and UPARCOS, Universidad Complutense de Madrid, 28040 Madrid, Spain}
\\
}
\maketitle
\begin{abstract}

We review our analysis of $\pi K$ scattering using forward dispersion relations. The method yields a set of simple parameterizations that are compatible with forward dispersion relations up to 1.6 GeV while still describing the data. Once the partial waves are obtained, we calculate the poles in the complex plane by means of Pad\'e approximants, thus avoiding a particular model for the pole parameterization. The resonances calculated below 1.8 GeV are the much debated scalar $\kappa$-meson, nowadays known as $K_0^*(800)$,
the scalar $K_0^*(1430)$, the $K^*(892)$ and $K_1^*(1410)$ vectors,  the spin-two
$K_2^*(1430)$ as well as the spin-three $K^*_3(1780)$.
\end{abstract}
  
\section{Introduction}

A reliable determination of strange resonances 
is by itself relevant for hadron spectroscopy and their own
 classification in multiplets, 
 as well as for our understanding of intermediate energy QCD and the low-energy regime through Chiral 
Perturbation Theory. In addition $\pi K$ scattering and
 the resonances that appear in it are also of interest
because most hadronic processes with net strangeness end up with at least a $\pi K$ pair
that contributes decisively to shape the whole amplitude through final 
state interactions. 

Very often the analyses of these resonances have been made 
in terms of crude models,
which make use of specific parameterizations like isobars, 
Breit--Wigner forms or modifications, which assume the existence of some simple background.
As a result, resonance 
parameters suffer a large model dependence or may
even be process dependent. Thus, the statistical uncertainties 
in the resonance parameters should be accompanied by 
systematic errors that are usually ignored. 

For the above reasons there is a growing interest
in methods based on analyticity properties 
to extract resonance pole parameters
from data in a given energy domain.
They are based on several approaches: conformal expansions
to exploit the maximum analyticity domain of the amplitude 
\cite{Cherry:2000ut},
Laurent \cite{Guo:2015daa}, Laurent-Pietarinen \cite{Svarc:2013laa} expansions,
 Pad\'e approximants \cite{Masjuan:2013jha,Pelaez:2016klv}, or the rigorous dispersive approach \cite{Ananthanarayan:2000ht}.
They all determine the pole position without assuming a particular 
model for the relation between the mass, width and residue.
In this sense they are model independent analytic continuations to the complex plane. 

These analytic methods require as input
some data description. It has been recently shown \cite{Pelaez:2016tgi} that in the case of $\pi K$ scattering data \cite{Estabrooks:1977xe}, which are the source for several determinations of strange resonances, 
they do not satisfy well Forward Dispersion Relations up to 1.8 GeV.
This means that in the process of extracting data by using models,
they have become in conflict with causality.
Nevertheless, in \cite{Pelaez:2016tgi}
the data were refitted 
constrained to satisfy those Forward Dispersion Relations and 
a careful systematic and statistical error analysis was provided.
In \cite{Pelaez:2016klv} we made use of the Pad\'e approximants method 
in order to extract the parameters of all resonances appearing in those waves.

In \cite{Pelaez:2016tgi} we used a set of fixed-t dispersion relations with $t=0$ so that we could implement this set of equations up to arbitrary energies in the real axis, providing a set of simple but powerful constraints for the fits. We considered two independent amplitudes, one symmetric and one anti-symmetric under the $s \leftrightarrow u$ exchange that cover the isospin basis $T^+(s)=(T^{1/2}(s)+2T^{3/2}(s))/3=T^{I_t=0}(s)/\sqrt{6}$ and $T^-(s)=(T^{1/2}(s)-T^{3/2}(s))/3=T^{I_t=1}(s)/2$. The symmetric has one subtraction and can be written as
\vspace{-0.15cm}
\begin{align}
&\re T^+(s)=T^+(s_{th})+\frac{(s-s_{th})}{\pi}\times \nonumber \\
&\times P\!\!\!\int^{\infty}_{s_{th}}\!\!\!\! ds'\!\left[\!\frac{\im T^+(s')}{(s'-s)(s'-s_{th})} 
-\frac{\im T^+(s')}{(s'+s-2\Sigma_{\pi K})(s'+s_{th}-2\Sigma_{\pi K})}\!\right],
\label{eq:FDRTsym}
\end{align}
\vspace{-0.1cm}
where $s_{th}=(m_\pi+m_K)^2$.
In contrast the anti-symmetric one does not require subtractions:
\vspace{-0.1cm}
\begin{equation}
\re T^-(s)=
\frac{(2s-2\Sigma_{\pi K})}{\pi}P\!\!\int^{\infty}_{s_{th}}\!\!\!\! ds'
\frac{\im T^-(s')}{(s'-s)(s'+s-2\Sigma_{\pi K})}.
\label{eq:FDRTan}
\end{equation}

We also included in our analysis 3 sum rules for threshold parameters (scattering lengths and slopes) in order to obtain the best possible result in this region, where there are no data.

\section{Method and results}

The first part of the calculation is to obtain a set of partial waves compatible with Eq. \eqref{eq:FDRTan}, however, the final result must also describe the data, at least qualitatively.
\vspace{-0.1cm}

\begin{figure}
\centering
\centerline{\includegraphics[width=0.42\linewidth]{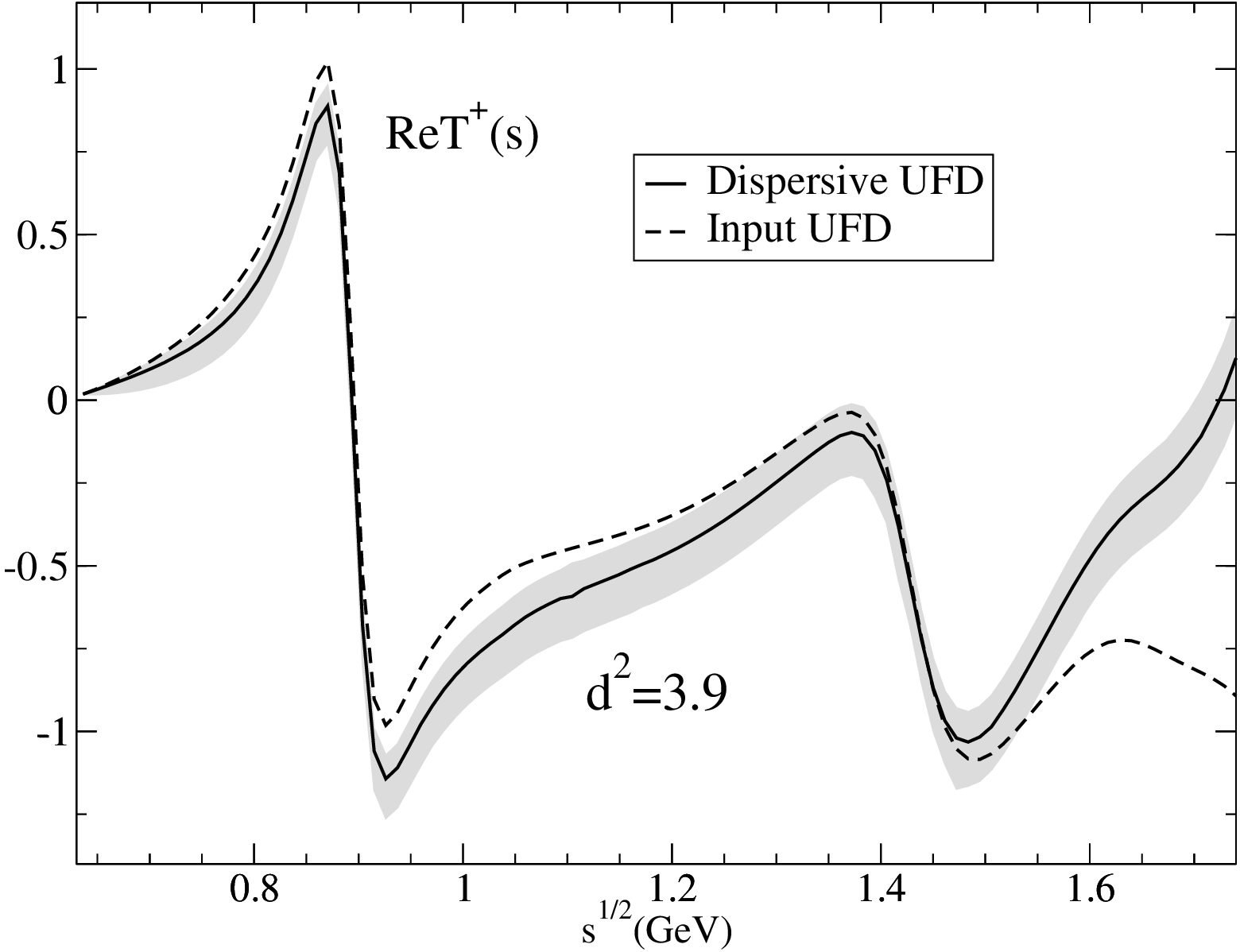}   \includegraphics[width=0.42\linewidth]{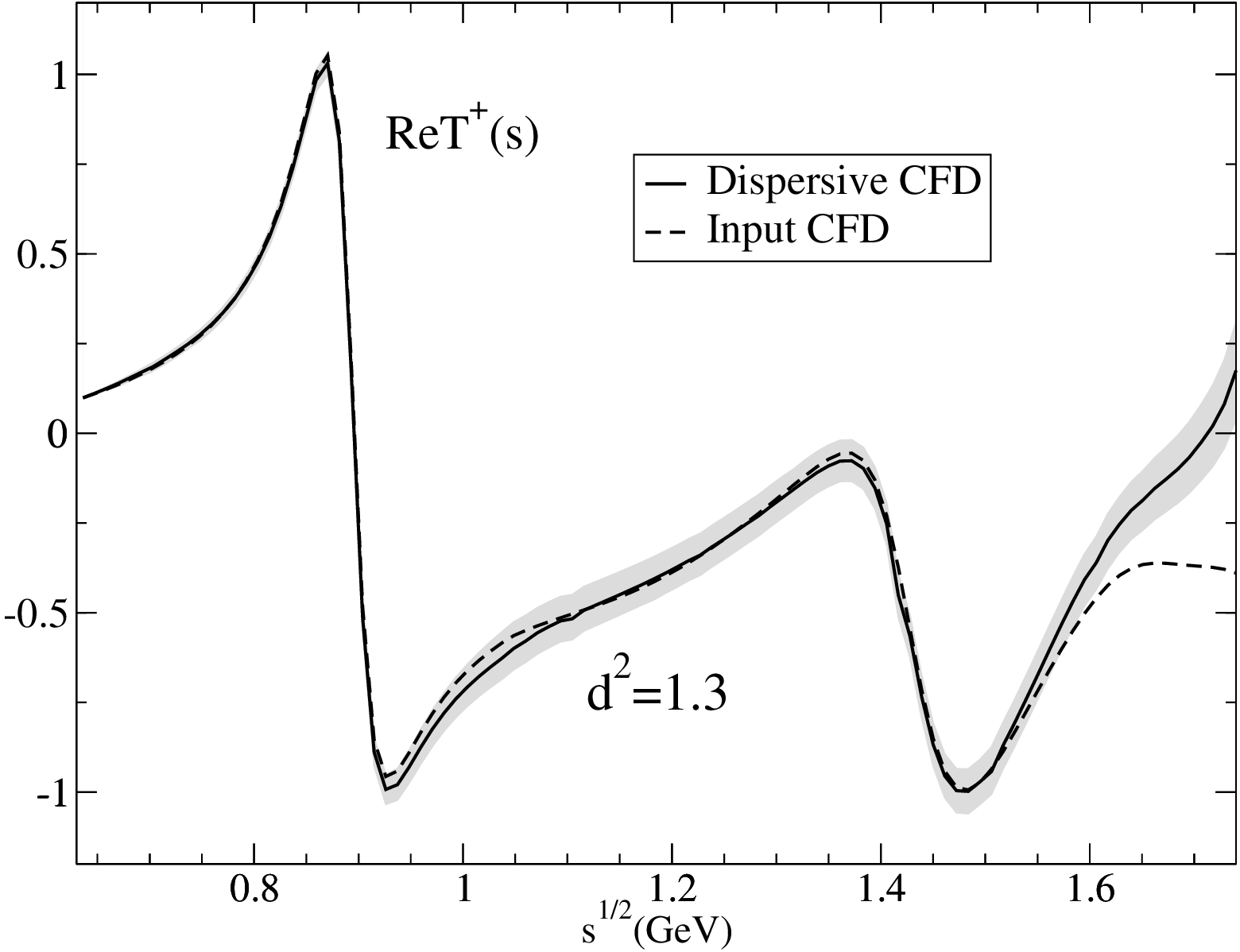}}
\vspace{0.15 cm}
\centerline{\includegraphics[width=0.42\linewidth]{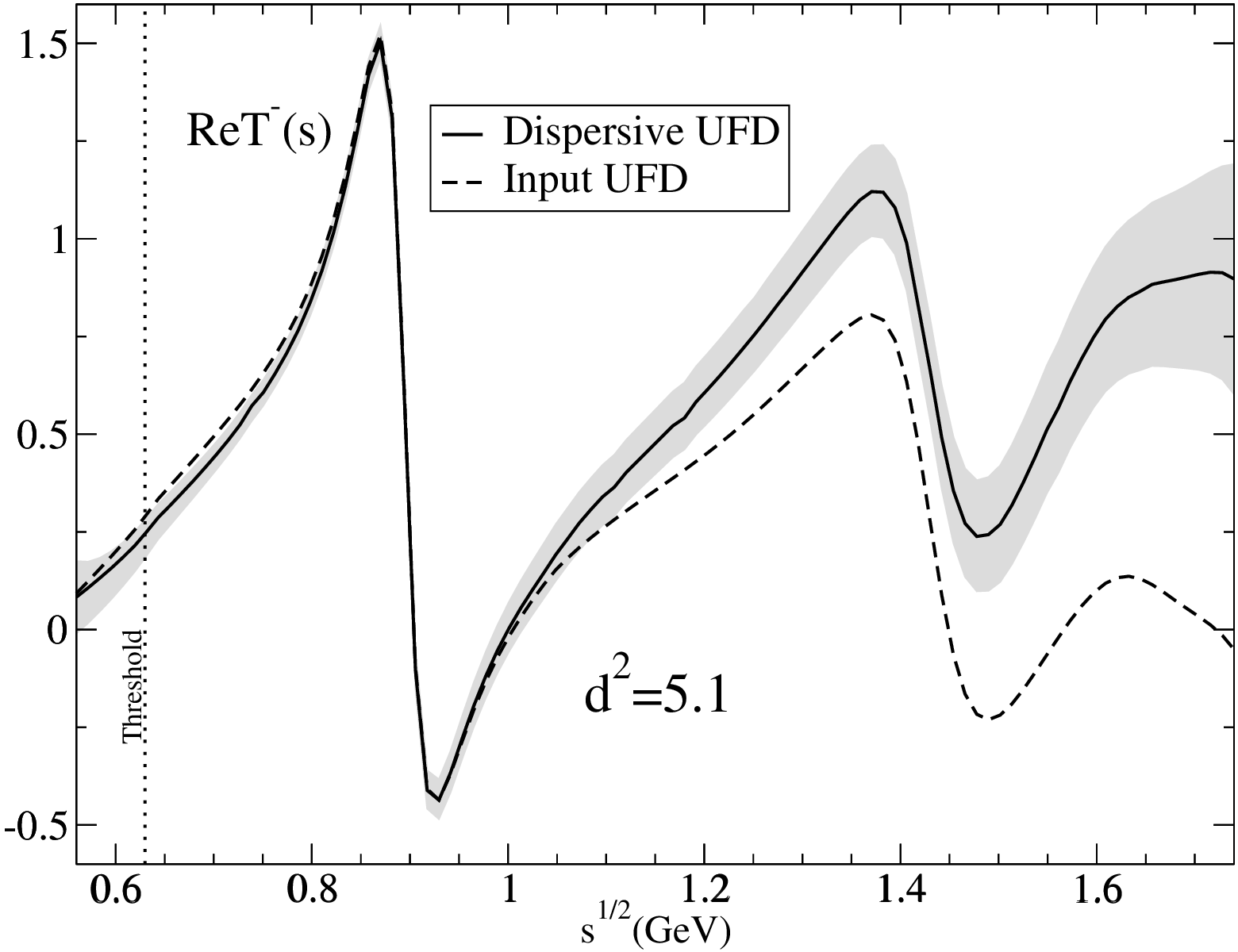} \includegraphics[width=0.42\linewidth]{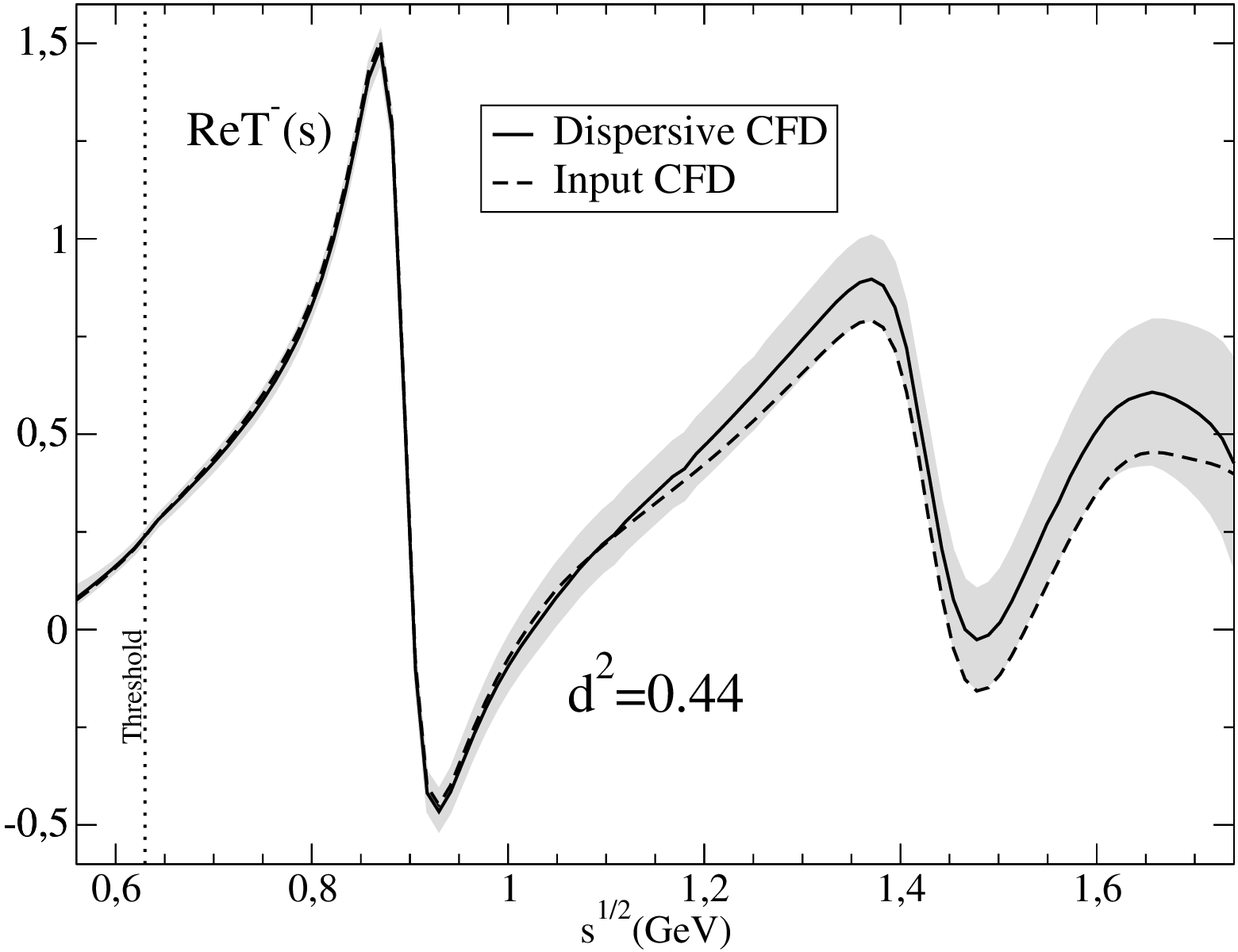}}
\vspace{-0.1 cm}
\caption{\rm \label{fig:FDR} 
Comparison between the input (fits)
and the output(FDRs) for the total amplitudes $T^+$ (top) and $T^-$(bottom).
The gray bands describe the uncertainty of the difference between the input and the output.
}
\end{figure}

In order to impose the FDRs we define a function $d_i$ as the difference between the input and the output of each dispersion relation at the energy point $s_i$, whose uncertainties are $\Delta d_i$. We thus define the average discrepancies
\begin{equation} 
d_{T\pm}^2=\frac{1}{N}\sum_{i=1}^{N} \left(\frac{d_i}{\Delta d_i}\right)_{T^\pm}^2.
\label{eq:distances}
\end{equation}
We also include a penalty function to ensure that the new solution still describes the data and then minimize the total function.

\begin{figure}
\centering
\centerline{\includegraphics[width=0.42\linewidth]{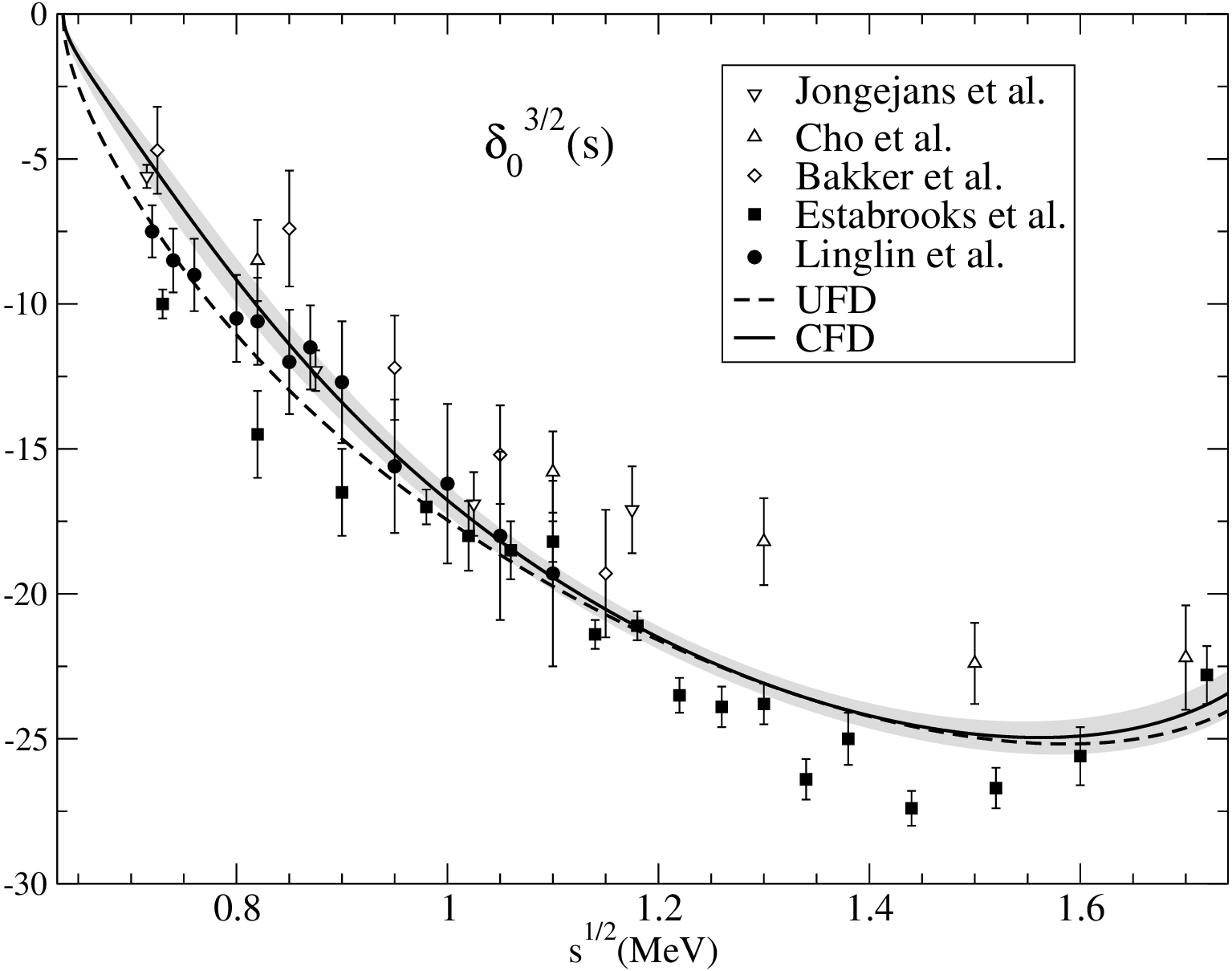} \includegraphics[width=0.5\linewidth]{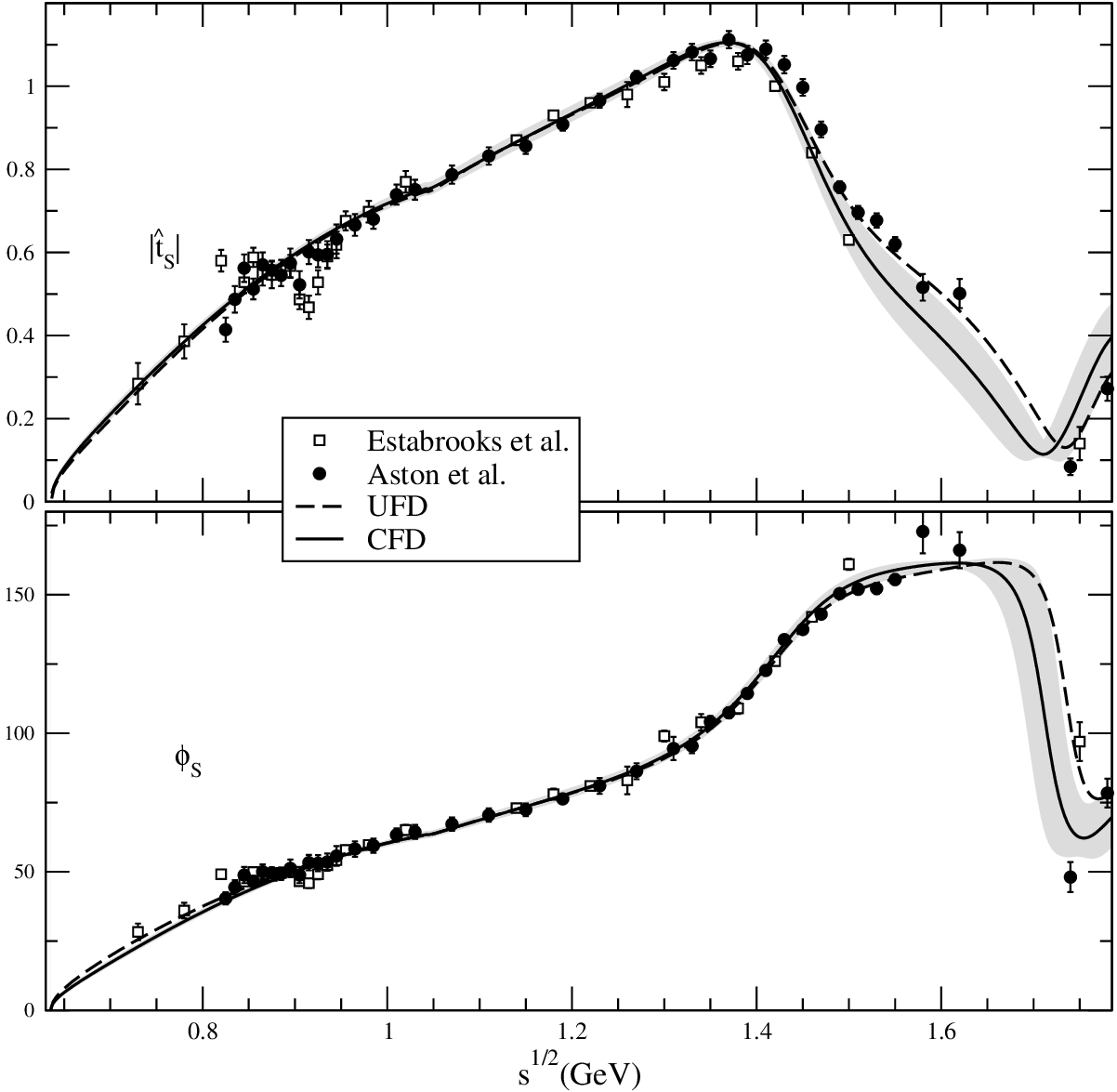}}
\caption{\rm \label{fig:CFD} 
Comparison between UFD and CFD fits for the scalar partial waves, where $|\hat{t}|$ stands for the modulus, $\delta$ for the phase shift and $\phi$ for the total phase of each partial wave. The gray bands cover the errors of the parameters for each fit.
}
\end{figure}

Fig.\ref{fig:FDR} shows the total amplitudes and the huge improvement between the UFD and the CFD parameterizations, in Fig.\ref{fig:CFD} we show the difference between the fits to the data and the final results for the scalar partial waves (region where the $\kappa$ exists). The scattering lengths obtained are compatible with some rigorous predictions and experimental determinations, reading $m_\pi a_0^{1/2}=0.22\pm 0.01$ and $m_\pi a_0^{3/2}=-0.054^{+0.010}_{-0.014}$.
Once we have obtained a set of equations that are compatible with the analytical requirements we can use the Pad\'e approximants to continue it to the complex plane. The $P^N_M(s,s_0)=Q_N(s,s_0)/R_M(s,s_0)$  Pad\'e approximant 
of a function $F(s)$ is a rational function that satisfies $P^N_M(s,s_0)=F(s)+O((s-s_0)^{M+N+1})$, 
with  $Q_N(s,s_0)$ and $R_M(s,s_0)$ polynomials in $s$ of order $N$ and $M$, respectively. 	In the case of one pole in the complex plane the formula reads
\vspace{-0.2cm}
\begin{equation}
P^N_1(s,s_0)=\sum^{N-1}_{k=0}{a_k(s-s_0)^k+\frac{a_N(s-s_0)^N}{1-\frac{a_{N+1}}{a_N}(s-s_0)}},
\end{equation}
\vspace{-0.2cm}
where the position and residue of the pole are
\begin{equation}
s_p^{N}=s_0+\frac{a_N}{a_{N+1}},~Z^{N}=-\frac{(a_N)^{N+2}}{(a_{N+1})^{N+1}}.
\label{eq:poleresidue}
\end{equation}
With this simple analytical continuation we can go to the next continuous Riemann sheet and find not only the elastic but also inelastic heavy resonances. We define the position of the pole as $\sqrt{s_p}=M-i\Gamma/2$, where the systematical errors of each pole are calculated using different parameterizations fulfilling FDRs and the statistical errors are estimated running a simple montecarlo for the parameters of each fit.

In the case of the $\kappa$ resonance, which is the lightest strange resonance (not confirmed according to the PDG), the calculation is compatible with the most rigorous dispersive result, showing the good agreement between both analytical methods. The result is $\sqrt{s_p}=(670\pm 18)-i(295\pm 28)$ MeV, while the result estimated by the PDG is $\sqrt{s_p}=(682\pm 29)-i(274\pm 12)$ MeV. The values obtained for the rest of the strange resonances appearing below 1.8 GeV are listed in Table \ref{tab:resonances}

\begin{figure}
\centering
\includegraphics[width=0.6\linewidth]{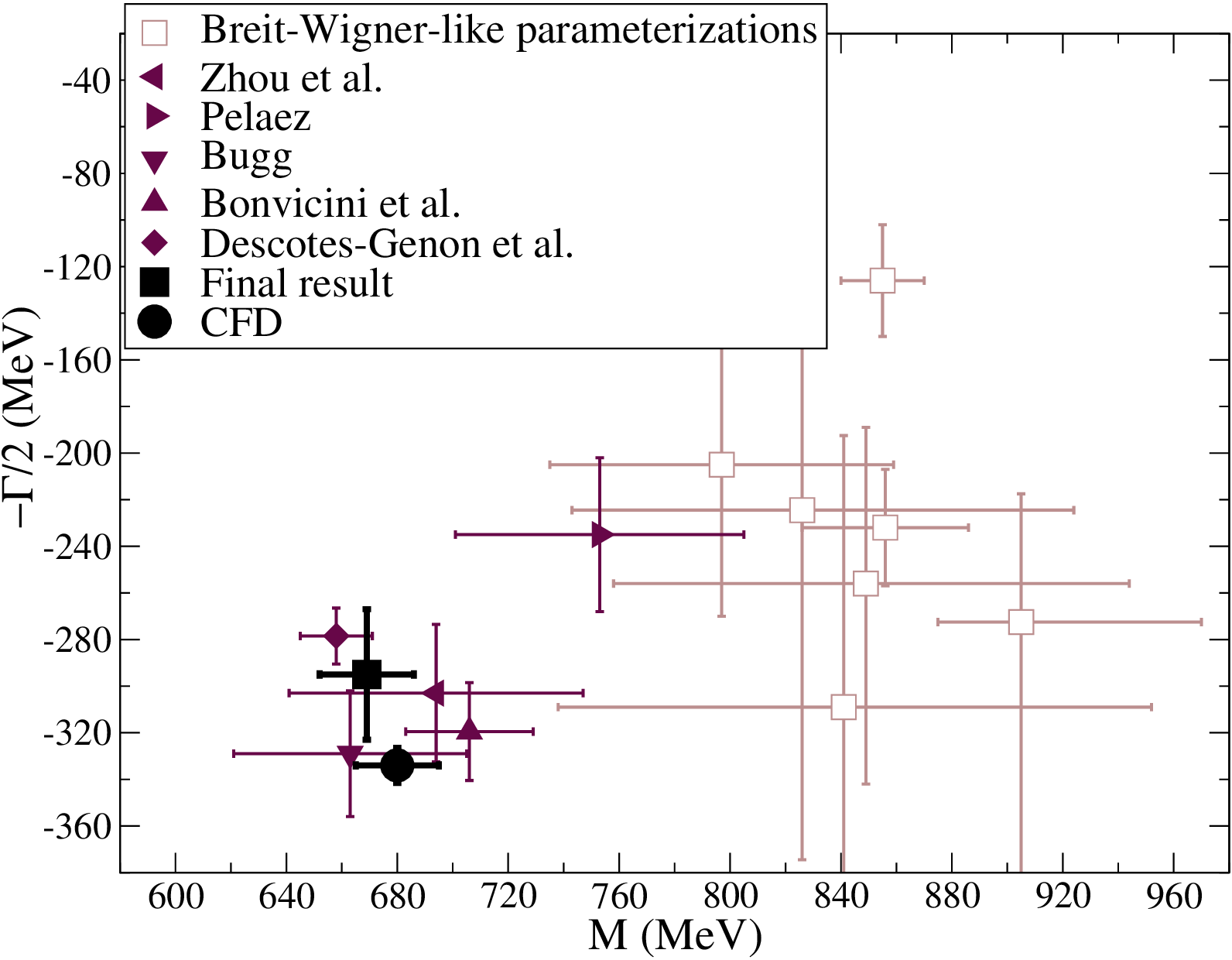}
\caption{\rm \label{fig:kappa} 
Final result for the $\kappa$ pole. Other references are taken from the RPP compilation \cite{PDG}.}
\end{figure}

\begin{table}[h] 
\caption{Resonance parameters.} 
\centering 
\begin{tabular}{c r r } 
\hline\hline  
Resonance & Mass (MeV) & Width (MeV) \\ 
\hline 
$K^*_0(1430)$  & 1431$\pm$ 6  & 110$\pm$ 19 \\
$K^*_1(892)$   & 892$\pm$ 1   & 29$\pm$1 \\
$K^*_1(1410)$  & 1368$\pm$ 38 & 106$^{+48}_{-59}$ \\
$K^*_2(1430)$  & 1424$\pm$ 4  & 66$\pm$ 2 \\
$K^*_3(1780)$  & 1754$\pm$ 13 & 119$\pm$ 14 \\
\hline 
\end{tabular} 
\label{tab:resonances} 
\end{table} 
\vspace{-0.5cm}

\section{Summary}

Fig.\ref{fig:FDR} shows that the CFD set satisfies really well the dispersion relations up to 1.6 GeV. Above that energy the differences between the input and the output require larger deviations from  data as it is shown in Fig.\ref{fig:CFD}.

Using the parameterizations obtained in \cite{Pelaez:2016tgi} we have calculated in \cite{Pelaez:2016klv} the parameters of the strange resonances appearing up to 1.8 GeV thanks to the method of the Pad\'e approximants. The values obtained for the parameters of the resonances are in agreement with other works in the PDG, although our approach is based on data analysis consistent with analyticity and makes use of a model independent method to extract the parameters, providing a realistic estimate of systematic uncertainties.

\section{Acknowledgments}
This work is supported by the Spanish Projects  FPA2014-53375-C2-2,
FPA2016-75654-C2-2-P and the group UPARCOS
and the Spanish Excellence network HADRONet FIS2014-57026-REDT.
A. Rodas would also like to acknowledge the financial support of 
the Universidad Complutense de Madrid through a predoctoral scholarship.

\end{document}